\documentclass{aa}
\usepackage{graphicx,epsfig}
\def\deg{^{\circ}}                                                  
\def\etal{{\it et al.}\thinspace}
\def\ie{{\it i.e.,}\thinspace}
\def\eg{{\it e.g.,}\thinspace}

\def\bei{\begin{itemize}}
\def\eei{\end{itemize}}
\def\bef{\begin{figure}}
\def\eef{\end{figure}}
\def\ben{\begin{enumerate}}
\def\een{\end{enumerate}}
\def\beq{\begin{equation}}
\def\eeq{\end{equation}}
\def\ber{\begin{eqnarray}}
\def\eer{\end{eqnarray}}
\def\p2{P_2}

\begin{document}\title{PSR B0809+74: Understanding Its Perplexing \\
 Subpulse-separation ($\p2$) Variations}

\author{Joanna M. Rankin\inst{1}\thanks{On leave from Physics Dept.,
  University of Vermont, Burlington, VT 05405 USA, email:
  joanna.rankin@uvm.edu} 
\and R. Ramachandran\inst{2} 
\and Svetlana A. Suleymanova\inst{3}}
\offprints{joanna.rankin@uvm.edu}

\institute{Sterrenkundig Instituut ``Anton Pannekoek'', NL-1098 SJ
  Amsterdam \email{jrankin@astro.uva.nl} 
\and Department of Astronomy, University of California, Berkeley, CA 94720 
    \email{ramach@astron.berkeley.edu} 
\and Pushchino Radio Astronomy Observatory, 142290 Pushchino, Russia         \email{suleym@prao.psn.ru} }

\abstract{The longitude separation between adjacent drifting subpulses, 
$\p2$, is roughly constant for many pulsars. It was then perplexing when 
pulsar B0809+74 was found to exhibit substantial variations in this measure, 
both with wavelength and with longitude position within the pulse window.  
We analyze these variations between 40 and 1400 MHz, and we show that
they stem primarily from the incoherent superposition of the two orthogonal 
modes of polarization.
}

\date{Received / Accepted }
\authorrunning{Rankin, Ramachandran \& Suleymanova}
\titlerunning{$P_2$ variations in PSR B0809+74}
\maketitle

\section{Introduction}
\label{intro}
PSR B0809+74 surely remains one of the most studied and influential 
``drifters'' in the pulsar literature. It exhibits all six of the fundamental 
pulsar phenomena: sub-pulse drifting, pulse nulling, profile mode 
switching, orthogonal emission modes, microstructure and 
``absorption''.  Its well known $\sim$11-period fluctuation feature 
was identified by Taylor, Jura \& Huguenin (1969) only seven
months after Drake \& Craft's (1968) discovery of the ``drifting''-subpulse 
phenomenon (in B1919+21 and 2016+28), but their poor resolution 
prevented the feature from being prominent. Vitkevich \& Shitov 
(1970) and Sutton \etal\ (1970) first exhibited the star's marvelously 
precise ``drifting'' pulse sequences (hereafter PSs) early the next 
year, and the latter paper introduced the now standard terminology 
of $P_1$, $P_2$ and $P_3$ for the pulsar rotation period, the 
subpulse-separation interval, and the driftband-separation 
period, respectively. Properties of these drifting subpulses have 
been investigated by many authors over the past three decades 
(\eg Cole 1970; Taylor \& Huguenin 1971; Backer \etal\ 1975). Its
polarization properties have also been studied in detail by various
investigators (Ramachandran et al. 2002; von Hoensbroech \& 
Xilouris 1997; Gould \& Lyne 1998).

A spectacular reported property of B0809+74's subpulse drift is 
the apparent variations of mean separation of subpulses ($P_2$) 
with longitude and frequency. Different (and often conflictual) 
values can be found in at least a dozen papers, and the reported 
frequency dependences, first in Taylor \etal\ (1975) and then in 
Bartel (1981) have influenced understanding of the properties of 
this pulsar greatly. According to Bartel (1981), the frequency 
dependence of $\p2$ is the same as that of the pulse width at 
various frequencies. He derives a frequency dependence of 
$\p2 \propto\nu^{-0.23}$. Similarly, its longitude-dependent 
variations have also been studied in the literature. According to 
van Leeuwen \etal\ (2002), the drift bands seen in this pulsar at 
328 MHz are not ``straight'', which implies that $\p2$ varies 
across the pulse profile. These results pose a serious problem 
for the geometrical interpretation of drifting subpulses. If one 
assumes that these drifting subpulses are a set of subbeams in 
the magnetosphere of a pulsar and that they drift around due 
possibly to $\vec{E}$$\times$$\vec{B}$ forces, then it is very difficult 
to reconcile these  observed longitude and frequency variations 
in subpulse spacing $\p2$ with such a geometrical standpoint.

\begin{figure*}
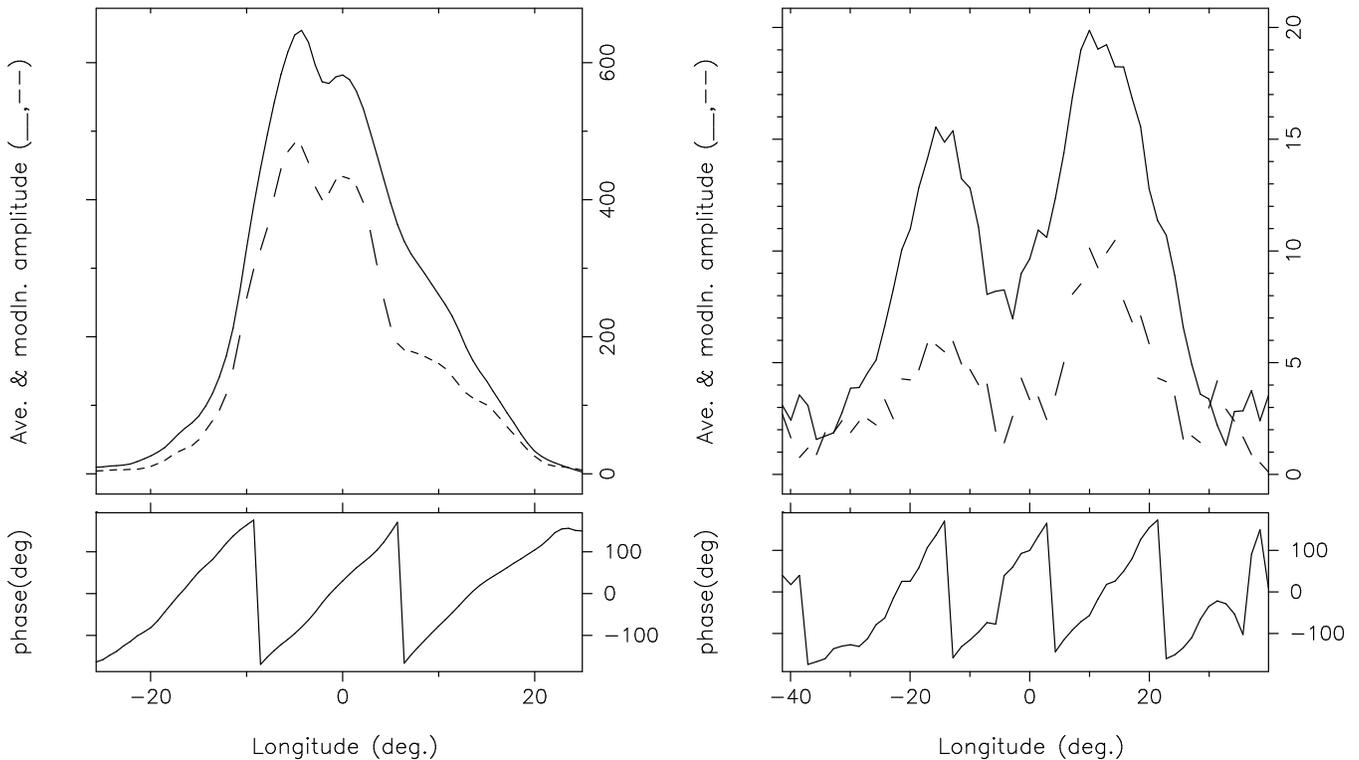

\begin{center}
\begin{tabular}{@{}lr@{}}
{\mbox{\epsfig{file=fig1a,height=8.7cm,angle=-90}}}&
{\mbox{\epsfig{file=fig1b.ps,height=8.7cm,angle=-90}}}\\
\end{tabular}
\end{center}
\caption{The total-power modulation amplitude and phase of 
B0809+74 at 112.7-MHz (left) and 40.9 MHz (right). The Stokes 
$I$ profile (solid) and modulation amplitude (dashed) are given 
in the upper panel; whereas the modulation phase is plotted in 
the lower panel. Note that here the phase rate is essentially 
constant over the body of the profile (except, as expected, in 
the very center of the 40.9-MHz profile), unlike that at higher 
frequencies.  These PRAO observations were recorded on 2 
December 2000 and 27 December 2003, respectively.
\label{fig8}}
\end{figure*}

Curiously enough, there appears to be little or no difficulty in
defining $P_2$ at low frequency. Between 81.5 and 151 MHz, six 
writers give values around 53$\pm$2 ms, or some 15.3$\deg$ of 
pulse longitude (Cole 1970; Sutton \etal\ 1970; Vitkevich \& Shitov 
1970; Page 1973; Bartel \etal\ 1981; Davies \etal\ 1984).  A key to 
understanding why this is so may follow simply from Davies \etal's 
discussion: They find that the driftbands at 102 MHz are ``essentially 
straight'', whereas those at 406 and 1412 MHz become increasingly 
curved.  Our own more recent observations confirm this conclusion 
as can be seen in the 112.7-MHz PRAO observation in Figure~\ref{fig8} 
(left).  Many of the reported $P_2$ values at around 400 MHz are 
only a little smaller than those above, apparently reflecting the fact 
that the driftbands curve in such a way as to parallel the low frequency 
behavior on the trailing edge of the profile.  This can be very clearly 
seen in Davies \etal's fig. 4.\footnote{The 40.9-MHz phase rate in 
Fig.~\ref{fig8}b implies a substantially larger $P_2$ value, some 65 
ms or 18$\deg$, apparently reflecting the overall larger longitude 
scale of the profiile.  We will discuss the implications of this value 
in a subsequent paper.}

Only at 21 cm does the literature report very different behaviors 
(\eg 29 ms, Bartel \etal\ 1981; 31 ms, Davies \etal\ 1984), and the 
reason for this can first be discerned in Wolszczan \etal's (1981) 
fig. 5a. Here we see that the drift-modulation phase exhibits a 
discontinuity near the profile peak amounting to about +180$\deg$---a 
circumstance which Edwards \& Stappers (2003; hereafter E\&S) 
as well as ourselves (not shown), have also confirmed.  Clearly, 
most early $P_2$ determinations at 21 cm were carried out in the 
vicinity of this drift-phase discontinuity under low signal-to-noise 
conditions in which this phase ``jump'' was not resolved, thus it is 
not at all surprising that very different values were obtained.  
Wolszczan \etal\ obtain much larger $P_2$ values, and if the 
180$\deg$ ``jump'' is roughly corrected for---that is, by estimating 
that 1.5 modulation cycles occur over about a 20$\deg$ longitude 
interval in their fig. 5a---we find that $P_2$ over the entire width of 
the profile may well be not far from the quoted 47 ms or 13.1$\deg$.  
This value, in turn, compares well with the 100-MHz value of 
15.3$\deg$, being only a little smaller (owing to the reduced overall 
scale of the profile) as might be expected---a conclusion also 
reached by Edwards \& Stappers (2003) who argue that the 
modulation-phase {\it rate} must be independent of frequency.  

The origin, however, of the phase ÒjumpÓ remains as yet unexplained.  
Why, perversely, should the phase rate---which is nearly linear at low 
frequency---first steepen in the centre of the profile at meter 
wavelengths and finally exhibit a 180$\deg$ discontinuity at 21cm?  
The reason, almost certainly, is modal polarization which we will try 
to demonstrate concretely below.  We must also stress that virtually 
every existing drift-modulation study of pulsar B0809+74, historical 
and more recent, was based entirely on the total power (Stokes $I$).  
We will return to this argument once we have first both discussed our 
observations and what is known of the star's modal polarization.

\section{Observations}
\label{sec-obsvn}
Observations used in our analyses below come from several different
sources. We make considerable further use below of the same remarkably
bright, 328-MHz polarized pulse sequence acquired using the Westerbork
Synthesis Radio Telescope (hereafter WSRT) on 2000 November 26 and
first studied by Ramachandran \etal\ (2002).  This observation has now 
been corrected for a recently determined instrumental effect that converted 
some linear into circular polarization (Edwards \& Stappers 2004; see their 
Appendix).  The 1.38-GHz observation was also made using the WSRT on 
10 January 2002 using an 80-MHz bandwidth, a 0.8192-ms sampling 
interval, and 256 effective channels across the passband.  The 112.7-MHz 
observations were made at the Pushchino Radio Astronomy Observatory 
(PRAO) using the BSA (Bolshaya Synfaznaya Antenna) telescope at 
112.7 MHz.\footnote{The instruments are fully described in the Pushchino 
Observatory website:http://www.prao.psn.ru} The signals from the linearly 
polarised array were fed to a radiometer with 128x20-kHz contiguous 
channels in order to measure the total intensity and its spectral variations 
across the passband.  The dedispersed pulses were referred to the 
frequency of the first channel---that is, to 112.67 MHz.  By observing a 
partially linearly polarised pulsar signal at adjacent frequencies the rotation 
measure can be obtained, and this in turn used to determine the linear 
polarization (Suleymanova, Volodin \& Shitov 1988) from the 
Faraday-rotation-induced, quasi-sinusoidal intensity modulation across 
the passband.  The time resolutions were 2.56 ms or 0.71$\deg$.  The 
observations were comprised of 555 pulse periods as limited by the BSA's 
beam.  The remarkable 41-MHz observation was acquired with PRAO's 
DKR-1000 instrument, using a 128x1.25-kHz radiometer and 5.12-ms 
integration.  The 27 December 2003 recording includes a nearly 
interference-free segment 947 pulses long, which was used for the present 
analysis.  Our analyses are based on high quality pulse sequences from 
which the null intervals have been removed, as indicated by the analyses 
and techniques developed by van Leeuwen \etal\ (2002).  Such a removal 
represents a first-order correction to the well known subpulse-phase near 
stasis across nulls. The effects cannot be completely removed, however, 
both because of higher-order effects and because some number of nulls, 
less than one $P_1$ in duration, will occur when the star beams in other 
directions.

\begin{figure*}
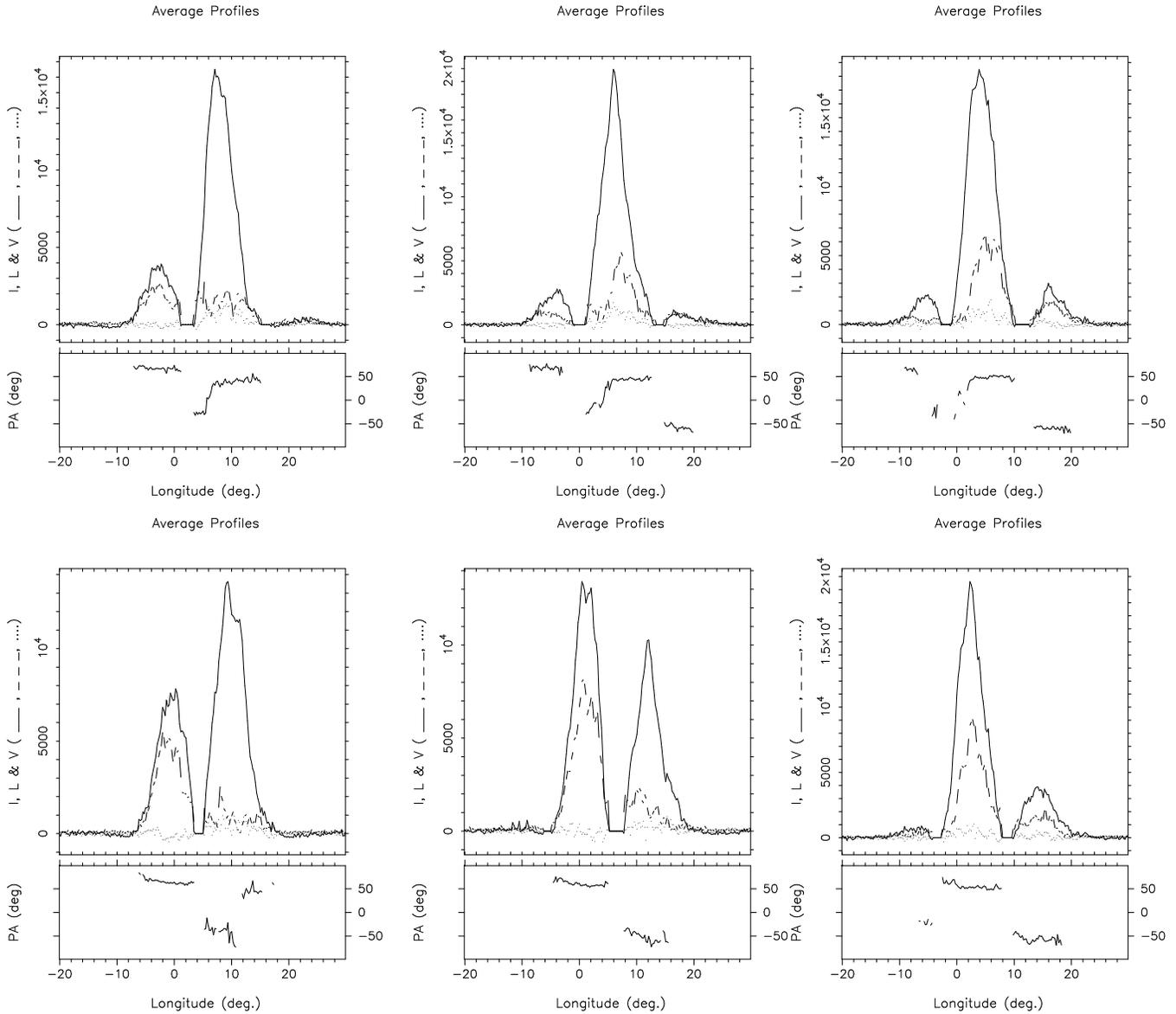

\begin{center}
\begin{tabular}{@{}lr@{}lr@{}}
{\mbox{\epsfig{file=fig2a.ps,height=5.8cm,angle=-90}}}&
{\mbox{\epsfig{file=fig2b.ps,height=5.8cm,angle=-90}}}&
{\mbox{\epsfig{file=fig2c.ps,height=5.8cm,angle=-90}}}\\
{\mbox{\epsfig{file=fig2d.ps,height=5.8cm,angle=-90}}}&
{\mbox{\epsfig{file=fig2e.ps,height=5.8cm,angle=-90}}}&
{\mbox{\epsfig{file=fig2f.ps,height=5.8cm,angle=-90}}}\\
\end{tabular}
\end{center}
\caption{``Cyclical'' clockwise composite of 6 partial polarisation 
profiles ({\it a,b,c,f,e,d,a, ...}) folded at 60$\deg$ phase intervals 
around the nominal 11-$P_1$ modulation cycle at 328 MHz. All were 
computed from the same section of this (recalibrated) observation, 
and two or three subpulses are seen which ``drift'' steadily from 
later to earlier phases about 1--2$\deg$ from plot to plot. The 
outside curves give the total power (Stokes parameter) $I$, 
the dashed ones the linear polarization $L$ (=$\sqrt{Q^2+U^2}$), 
the dotted ones the negligible circular polarization $V$, and 
the others the polarization angle $\chi$ (=$0.5 \tan^{-1}U/Q$). Note
that the earliest subpulses (here folded to an average) have a
positive PA, while the latest have a negative PA. The broader 
central subpulses exhibit the most interesting behavior, often 
showing adjacent regions of modal polarization and substantial 
depolarization. Panels {\it e} clockwise around to {\it b} exhibit 
this behavior very clearly. Note the depolarization in the  first 
two as well as the PA ``jumps'' in the last three. Can it be doubted 
that modal polarization effects play a major role in the pulsar's 
strange longitude variations in subpulse width and spacing $P_2$?}
\label{fig3}
\end{figure*}

\section{Properties of orthogonal modes in PSR B0809+74}

A very important property of pulsar subpulses is that they do 
not have any memory of their polarization state.  During their 
drift  though our sightline, as we see in all pulsars that exhibit 
drifting-subpulse behaviour, their polarization state changes 
continuously and appears to depend entirely on their location 
within the ``pulse window'' or profile. This can be seen very 
clearly for pulsar B0809+74 at 328 MHz in the colour display 
given by Ramachandran \etal\ (2002) as their figure 4, where 
the two polarization modes have about equal strength, but vary 
dramatically---and very systematically---in relative intensity 
throughout the profile.\footnote{The circular polarization in this 
display should be discounted, and the fractional linear taken 
as being about 50\% greater per the recent WSRT recalibration 
by Edwards \& Stappers (2004).  The configurations of modal 
polarization, however, remain nearly unaltered.}  The behaviour 
is also demonstrated, using mode-segregation methods, for 
B0809+74 as well as other bright pulsars (see fig. 3 of Rankin 
\& Ramachandran 2003).  

Figure~\ref{fig3} provides an average, but accurate means 
of assessing the polarization characteristics of the 
``drifting''-subpulse pattern.  Its six panels give partial 
profiles corresponding to successively increasing intervals 
of phase within the star's overall $\sim$11-$P_1$ modulation 
cycle.  We thus see in a ``cyclical'' ({\it abc/fed/a}) order polarized 
subpulses appear on the the trailing edge of the total-power 
profile (at longitudes near 20$\deg$) and then drift progressively 
toward earlier longitudes from panel to panel.  The sensitivity 
of the analysis is such that the average subpulse position and 
properties can be traced through nearly three rotations around 
the plots, representing an interval in longitude of thrice the 
subpulse spacing---or, what is the same, that we generally 
see three subpulses in each plot (or individual pulse), with 
one barely appearing on the extreme trailing edge of panel 
{\it a} and another close to disappearing on the extreme 
leading edge in panel {\it f} or even {\it e}.\footnote{Manchester 
\etal's (1975) fig. 7 also clearly shows this progress of modal 
polarization effects along the driftband, but not in a form 
where the relationships between adjacent subpulses can 
be discerned.  }  

\begin{figure}
\begin{center}
\epsfig{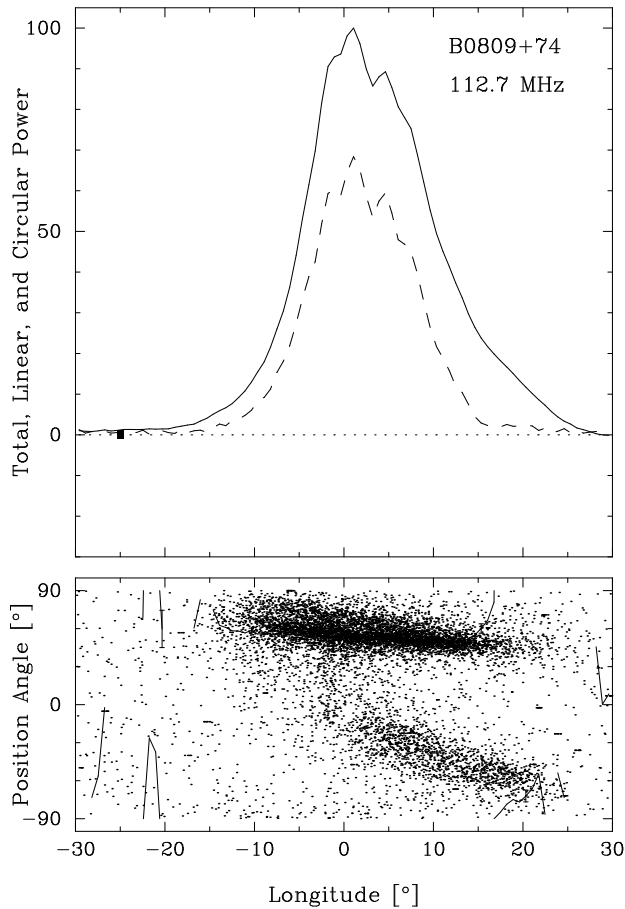}
\end{center}
\caption{Polarization histogram of a 112.7-MHz PRAO observation 
of 6 February 2000. The solid and the ``dashed'' lines give total power 
and linearly polarized average profiles. A high degree of linear
polarization is seen at this frequency, as one polarization mode 
seems to dominate. The linear polarization PA is given in the bottom 
panel as a function of pulse longitude. The longitude reference 
here is arbitrary, and the circular polarization was not measured.}
\label{fig9}
\end{figure}

A remarkable aspect of these subpulses is that they are so 
obviously polarized in a modal manner.  Subpulses appearing 
to the left of the diagrams are substantially linearly polarized 
and have negative PA values; whereas, subpulses that are 
waning at the right of the diagrams also have a fairly large 
linear polarization with positive PA values.  (Here the modal 
PA tracks fall conveniently about the PA origin in a roughly 
symmetrical manner; their individual behavior is very clear, 
for instance, in panel {\it b}).  What is arresting, however, is the 
behavior in between these extremes: the (average) subpulses 
peaking in the range between about 10 and 4$\deg$ longitude
are both depolarized and show a sharp modal transition---and 
note that the modal depolarization and PA ``jump'' marking 
this transition moves in an orderly manner from the trailing edge 
of the following subpulse in panel {\it e}, to the center of the 
bright subpulse in {\it d}, and then to its leading edge in panels 
{\it a} and {\it b}.  

The displays of Fig.~\ref{fig3} compliment what we can see 
in Ramachandran \etal's fig. 4.  There, the driftbands can be 
traced in total power (Stokes $I$) over almost as great a longitude 
interval, but the polarization can be delineated only over a 
span somewhat greater than $P_2$.  However, in the region near 
the 328-MHz profile peak, the colour pulse-sequence (hereafter 
PS) display provides better resolution of the polarization changes.  
(Note that the longitude scale of this PS display is not the same 
as that of Fig.~\ref{fig3}.)  Also, note that the fractional linear 
polarization and angle are remarkably consistent over the 
entire 200 PS displayed; whether one mode or the other is 
active (red vs. green coded angles in the 3rd column) $L/I$ is 
about 60\% (with the 50\% calibration enhancement mentioned 
above).  Finally, note that the modal ``tracks'' in positive 
(green-cyan) PAs are not parallel to those in total power; 
these have a somewhat earler apparent ``driftrate'' and are 
encountered at earlier longitudes; whereas the other mode 
(red-magenta PAs) is encountered later and is more parallel 
to the total-power drift ``track''.  Indeed, the overall 
effect of the latter is to broaden the driftband at and before 
the average profile peak.  

While these effects can only be estimated when looking at the 
polarized PS display, Fig.~\ref{fig3} exhibits their effect more 
clearly.  The (average) subpulses near the peak of the profile 
exhibit adjacent regions of modal power and are thereby 
somewhat broader in width.  This is close to what Davies \etal\ 
(1984) observed at their nearby frequency of 406 MHz as seen 
in their fig. 5 (though a different behavior was observed at 102 
MHz, which we will discuss below).  A number of different 
geometrical and other effects contribute to the varying subpulse 
width and spacing as a function of longitude, but we cannot 
doubt that these systematic modal polarization transitions are 
a major contributor.  

The orthogonal modes in B0809+74 also seem to have a 
different behaviour at different radio frequencies. For instance, 
at 112.7 MHz, the polarized emission is more or less completely 
dominated by one of the modes across the full width of the profile, 
as shown in Fig. \ref{fig9}. However, this is not the case at higher 
frequencies.  At 328 MHz (see Fig. \ref{fig3} and fig. 1 of 
Ramachandran \etal\ 2002\footnote{Note that we now know 
that the linear power is about 50\% greater than as shown in 
this figure because of the recalibration mentioned above.}), 
both the modes are clearly present, although with varying 
strengths across the profile. In fact, for a limited longitude range 
around the peak of the average pulse, strong subpulses 
dominated by both modes occur.  However, one mode 
dominates most subpulses in the leading part of the profile, 
and the other mode dominates in the trailing part. The 
orthogonal PA ``jump'' in the average profile occurs somewhere 
around the peak of the profile, but the low linear polarization 
throughout the profile shows that modal depolarization is a 
factor everywhere. This sort of behavior is seen clearly in the 
early polarimetry of Lyne \etal\ (1971) as well as in the 234- 
and 606-MHz profiles of Gould \& Lyne (1998).\footnote{The 
Gould \& Lyne (1998) 410-MHz profile seems exceptional in 
this regard.  Perhaps the star exhibits some modal variability, 
but both our own observations as well as those of Manchester 
(1971) around this frequency confirm the above behaviour.}

At higher frequencies, a different and quite unusual phenomenon 
appears, where the leading edge of the profile becomes almost 
fully linearly polarized. This can be seen at and above 925 MHz 
in Gould \& Lyne and is very finely exhibited in the 1.41- and 
1.71-GHz polarimetry of von Hoensbroech \& Xilouris (1997). 
Nearly complete linear polarization is virtually unknown among 
conal single ({\bf S}$_d$) stars, and remarkably suggests (a) that 
only one polarization mode is observed in this longitude range, 
and (b) that it is almost fully linearly polarized!  Once, then, we 
have gazed on this splendid 1.41-GHz profile, in which the entire 
uni-modal leading edge up the the 75\% power point is fully 
polarized and then noted the precipitous falloff in linear 
polarization to virtual depolarization on the trailing edge---a 
region where the two modes must be active with a depolarizing
effect---can we doubt that there will be a phase boundary of 
some sort between these two contrasting modal behaviors?  
---And, this phase boundary falls precisely where it must: 
between the peak and the trailing 3-db point of the linear 
power---or just before the PA excursions in the above profiles.

\section{PSR B0809+74's $P_2$ Variations}
\label{sec-p2var}
Careful perusal of our Fig.~\ref{fig3} above demonstrates that 
$P_2$ must vary as the modal configuration of the mean 
subpulses varies with longitude across the profile.  A steepening 
will then certainly occur near the center of the total intensity 
profile where fully double-moded subpulses must be fitted 
(crowded) into the driftbands.  Apparently, the somewhat 
simpler situation at low frequency arises because one mode 
dominates across most of the profile, so that their joint effect 
changes little with longitude.   That this is so can be seen in the
112.7-MHz polarization histogram shown in Figure~\ref{fig9}; note 
both the large fractional linear polarization and dominance of a 
single polarization mode across most of the profile. 

We have, however, not yet directly demonstrated in any detail 
that B0809+74's $P_2$ anomalies are due to modal polarization 
effects.  Surely, we should expect that virtually any sightline 
traverse along the edge of a conal beam will encounter systems 
of modal ``beamlets'', and we have shown (Ramachandran \etal's 
fig. 4) that the observed ``beamlet'' structure varies with longitude 
across the profile, but it remains to see what specific modal effects 
caused the historical difficulties that many investigators have had 
in determining the star's $P_2$, especially at around 1400 MHz.

\begin{figure}
\begin{center}
\epsfig{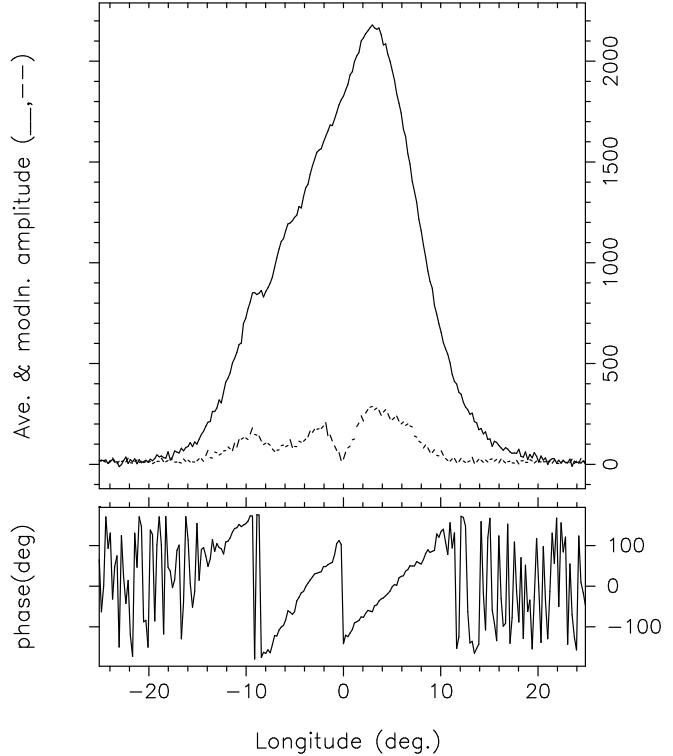}
\end{center}
\caption{Average profile (solid line), modulation amplitude (dotted
line) and modulation phase (bottom panel) at 1.38 GHz. The phase  
rate here varies a good deal across the profile and exhibits the
discontinuity near the longitude origin first resolved by Wolszcan
\etal. Note, however, that the phase rate on the leading and trailing 
edges of the profile is not far different than that seen at the lower 
frequencies.  This WSRT observation was recorded on 10 January 
2002.
\label{fig10}}
\end{figure}

First, let us examine in our own observations what others may 
have seen at 1.4 GHz. Figure~\ref{fig10} gives a total-power 
average profile for the pulsar at 1380 MHz and shows how the 
modulation phase varies across it. The modulation-feature width 
here is several FFT bins wide, so that only about 10\% of the 
power falls at the fluctuation frequency in question. Here, however, 
we see the usual phase rate---corresponding to a $P_2$ of about 
13$\deg$---under the trailing half of the profile (and the adjacent 
frequency bins behave similarly).  Note that the phase rate on 
the extreme leading edge of the profile is near this value, but that 
it steepens rapidly and then exhibits a large discontinuity near 
the longitude origin.  This behavior is very similar to that first 
reported by Wolszcan \etal\ (1981: fig 5a) and more recently by 
Edwards \& Stappers (2003: fig. 2b), but it is very unlike anything 
we have thus far encountered at lower frequencies.

In order to explore the polarization structure of the ``drift''-modulation, 
we have computed $I$$Q'$$V$ Stokes sequences in which nearly 
all of the linearly polarized power is rotated into a single Stokes 
parameter $Q'$. To the extent that the polarization modes are 
orthogonal, they can be represented simply in terms of positive or 
negative contributions to $Q'$---and if $U'$ is then found to contain 
only a negligible level of noise-like power, this procedure is well 
justified. $Q'$ then has the character of $V$ in representing the 
linearly polarized power along a principal axis of the Poincar\'{e} 
sphere---and given that this power is modal, the two modes will 
have opposite signs.

We have computed this $I$$Q'$$V$ sequence for our WSRT 
observations at 328 and 1380 MHz. In both cases the nulls had 
already been identified and removed, so that large portions of each 
null-removed sequence exhibited a very high-Q modulation feature.  
We tested the efficacy of this procedure by examining the Stokes-$U'$ 
PSs, and in both cases we found that so little modulation power 
remained in them that no fluctuation feature corresponding to the usual 
11-$P_1$ $P_3$ could be discerned. Virtually all of the modal 
fluctuation power was then represented by $Q'$.

\begin{figure}
\begin{center}
\epsfig{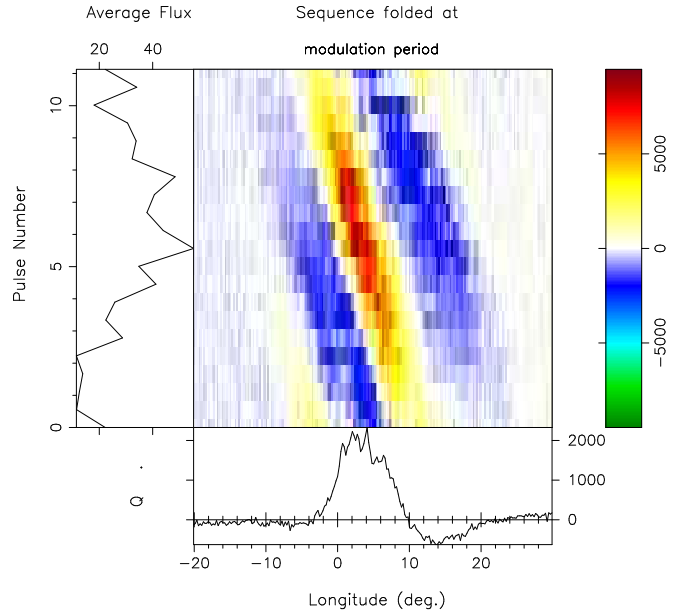}
\end{center}
\caption{Color intensity-coded diagram showing the distribution of
linearly polarized fluctuation power as a function of longitude over
the $11\times P_1$ modulation cycle at 328 MHz. One polarization mode is
positive (yellow-orange-red) and the other negative (blue-cyan-green)
in this representation. Note that the average $Q'$ power (left-hand
side panel) is quite small, reflecting the nearly complete linear
depolarization at this frequency; whereas, the bottom panel shows that
one mode (that here seen as positive) dominates in the early part of
the profile and the other the later. See text for details.}
\label{fig11}
\end{figure}

Figure~\ref{fig11} shows the result of folding $Q'$ over the precisely 
determined 11-$P_1$ modulation cycle at 328 MHz. The colored 
driftbands represent the behavior of the polarized modal power. It 
is immediately clear from the main panel that this power is not at 
all symmetrically distributed, as the positive (yellow-red) driftband 
spacing (\ie $P_2$) is markedly narrower than that of the negative 
(blue-green) mode. Also we can see that if integrated along each 
driftband, the negative mode exhibits a fairly flat power profile, 
while that of the positive mode is peaked and slightly skewed to the 
left. These are then the linear polarization characteristics which 
greatly complicate the $P_2$ determination in total power. One might 
measure a fairly consistent (but distinct) value of $P_2$ for each 
polarization mode across the entire pulse profile at 328 MHz; however, 
when only the intensity of these modal contributions is measured, 
there is no way to discern how complex and non-linear is their joint 
effect. Note, for instance, how the $Q'$ power varies over the 
modulation cycle in the left-hand panel and how it is distributed 
in longitude in the lower one.

\begin{figure}
\begin{center}
\epsfig{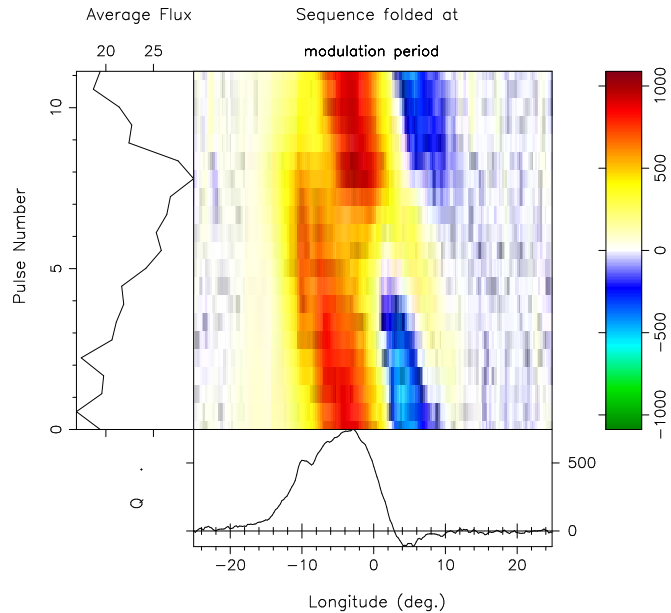}
\end{center}
\caption{Color intensity-coded diagram showing $Q'$ at 1.38 GHz 
folded over the $11\times P_1$ modulation cycle as in Fig.~\ref{fig11}. 
Notice that here the profile polarization is slightly positive 
(red-orange-yellow) over the entire modulation cycle (left-hand panel) 
but that this dominance is complete only during the leading part of 
the profile (see bottom panel). In the main panel we see that the 
subpulse modulation under the first part of the profile is produced 
entirely by the ``positive'' mode. The negative mode only appears 
after the longitude origin at just the point where $Q'$ decreases 
sharply. There is indeed a good deal of the negative-mode (blue-cyan-green) 
fluctuation power here as well, and this results in the nearly zero 
aggregate linear power. The longitude interval between the bright 
red and blue-cyan bands is only some 9$\deg$, which is just what 
would be measured in the total power. However, the phase rate 
of each band is compatible with a $P_2$ of some 13$\deg$ on the 
profile edges where it is dominant, as we have seen above. What 
this diagram shows most remarkably, however, is the way that the 
negative driftband ends surrounded by positive drift power and is 
then continued, almost precisely in line, by the positive drift band.}
\label{fig12}
\end{figure}

Using this same method, we can approach the questions which 
prompted this discussion: ($i$) why is $P_2$ so difficult to determine 
at 1.4 GHz, and ($ii$) what causes the modulation-phase ``jump'' near 
the center of the profile?  Both Figures~\ref{fig12} \& \ref{fig11} are 
produced by folding the respective time sequence precisely at $P_3$. 
Notice how differently the modal polarization power behaves at 
1380 MHz when compared with to 328 MHz in the previous figure. 
True, the leading edge of the profile is modulated by one bright mode, 
and the other appears more active within the later part of the profile. 
We can see clearly in the bottom panel that the two modes have 
about equal intensity ($Q'$ about zero) at positive longitudes---and 
this in strong contrast to the early part of the profile.

We saw at 328 MHz that the driftband phase cannot be linear across 
the profile, because subpulses with both polarization modes must be
accommodated in an environment where the drift involves only a single
mode on the profile wings.  Here, at 1.38 GHz we see a very different
behaviour: subpulses in the leading half of the profile exhibit only a
single polarization mode, whereas those in the trailing half reflect 
about equal contributions of both modes.  At the lower frequency, the 
two modes lie immediately adjacent to each other along modal subbeam 
``tracks'' only a few degrees wide, apparently assuming their positions 
along the driftband with such a precision that a line of modal 
depolarization is hardly discernible in Fig.~\ref{fig11}. Indeed, this 
is just what we see in the PS polarization display in Ramachandran 
\etal's (2002) fig. 4. 
 
At 1.4 GHz, however, we see a very conspicuous region near the centers
of both the profile and the driftband, where the driftband appears to
cease and then restart.  This represents quite a different behaviour
than just seen at lower frequency, where the modal driftbands parallel
each other for a large portion of the modulation cycle.  Here there
seems to be some modal depolarization {\it along} the overall driftband, 
as might happen, for instance, if there is some longitude irregularity in
the modal subbeam longitude positions.  Indeed, the width of the
driftband at 1.38 GHz does appear to be somewhat larger than that at
328 MHz, particularly under the first half of the profile.

In this rather complex context, we can well be less surprised that
multiple experts have, over the years, found this pulsar's $P_2$ so
difficult to measure around 21 cm. Upon looking at Figure~\ref{fig12} 
in more detail, though, we see that while the prominent modal 
driftbands on the leading and trailing edges of the profile have a 
drift rate that is compatible with that seen at lower frequencies, 
the modal driftband spacing in longitude is far less, about 9$\deg$, 
which indeed is one of the values often reported for $P_2$ at 1.4 
GHz. Here we can see the consequences of this modal complexity 
fairly clearly, but when studied in total power, as has heretofore 
uniformly been the case, such artefacts of the polarization-modal 
modulation remain impossible to discern and decipher.

Similarly, we can see here that the drift phase must change in the
center of the profile at 1.4 GHz for almost exactly the same reason as
the sign of Stokes parameter $Q'$ changes. The initial part of this
drift-phase trajectory is seen near the center of the diagram, where 
leftward-drifting subpulses develop comparable amounts of leading 
negative and trailing positive modal polarization.  A drift-phase 
discontinuity must then occur just were the negative part of these 
symmetrically modally polarized subpulses ceases---that is, just 
where the positive mode suddently becomes dominant in the lower 
panel.  Most significantly, however, Fig.~\ref{fig12} shows us that 
the predominantly negative driftband trajectory established in the 
latter part of the profile first broadens, wanes and is then picked 
up again, almost at the same driftband phase, by the opposite 
polarization mode.  

Or, said differently, Fig.~\ref{fig12} shows us that while the negative 
(blue-cyan-green) modal driftband in the latter part of the profile 
appears to extrapolate linearly to the positive (yellow-orange-red) 
one that continues in the earlier part of the profile, subpulses that 
are observed in total power will exhibit a modulation-phase 
``jump'' at the modal boundary.  This ``jump'' simply reflects the 
circumstance that the subpulses drifting into the middle of the 
profile from the trailing edge are polarimetrically bi-modal, 
whereas those in the earlier part of the profile exhibit only one 
mode.  A modulation-phase step is then required across this 
modal drift discontinuity at about 0$\deg$ longitude.  Some 
non-linearity or discontinuity must ``mark the spot''.   Indeed, 
such a ``jump'' can only occur because increasingly depolarized 
but underly subpulses, on average, carry the phase up to the point 
of the ``jump'' on both sides. Probably, this means that subpulses 
associated with both the bi- and unipolarizationmodal driftbands 
occupy the ``jump'' region in successive pulse, giving rise to the 
reports of ``confused'' ``drifting'' in this central region.

This is a remarkable result.  We thus now understand in an
analytical sense how it is that the large phase ``jumps'' occur in 
the 21-cm total-power driftbands of B0809+74---and perhaps 
some other stars as well.  Edwards \& Stappers (2003) have 
argued the such modulation-phase discontinuities can be 
produced by the superposition of out-of-phase modulation 
patterns, and here we see that such patterns can be generated 
by modal polarization.  We still have much to learn, however,  
about what modal ``beamlet'' configurations and/or sightline 
geometries can give rise to such a counter-intuitive modulation-folded 
polarization pattern.  This result does demonstrate for us that, modal 
polarization effects apart, we can still think of the subbeams as 
passing our sightline with a relatively constant and basically
geometrically-determined value of $P_2$---or, what is the same, our
fundamental cartoon conception of the subbeam configuration as a
``carousel'' survives. On this basis we can proceed much more
confidently in our further attempts to determine the full subbeam
configuration and circulation time.

\section{Summary and Discussion}
We have attempted in this paper to understand the character and 
causes of the (in)famous variations of subpulse spacing ($\p2$) in 
pulsar B0809+74.  Some dozen published studies report both that 
the subpulse spacing changes measurably with longitude at a 
given frequency and that significantly different values are obtained 
at different frequencies.  While this puzzling issue has lain dormant 
for several decades, it is essential to fully and accurately characterize 
the situation.   B0809+74 is usually quoted as exemplifying such 
effects, and recent studies are beginning to identify less dramatic 
but still important such variations in other pulsars.  If, in fact, we are 
to take the reported $\p2$ variations in B0809+74 as literally true, 
then it is very difficult to understand how the usual rotating-subbeam 
``carousel'' interpretation of subpulse drift could be retained.  

We find that B0809+74's $\p2$ variations are primarily artefacts of 
modal polarization in total-power analyses. This is one of the many 
effects introduced by mixing of the quasi-orthogonal modes. We thus 
clearly demonstrate that the use of total-power PSs for pulse-modulation 
studies will often lead to confusing results due to polarization-mode 
mixing, just as they are well known to do when averaged to produce 
polarized profiles.  Specifically, we find that---
\begin{itemize}\item The near constancy of $\p2$ at around 100 MHz and below is 
due to the circumstance that one polarization mode is dominant.  
This results in the ``straight driftbands'' observed by earlier workers.  \item Significant curvature is seen in the pulsar's driftbands at 328 
MHz, and this can be traced to the overlapping of the polarization 
modes over nearly the full width of the profile.  The ``straighter'' 
driftbands on the leading and trailing edges of the profile occur in 
regions where on mode is dominant. Conversely, the most curved 
section of the driftband is found near the longitude of the profile peak 
where both modes ``crowded'' into the same longitude interval at 
comparable intensities---and, of course, this is just where we find 
the profile most depolarized.
\item At 21 cms nearly complete linear polarization is observed in 
the leading half of the profile---this in strong contrast to most other 
conal single stars which exhibit highly depolarized profiles.  This 
suggests (a) that this polarization mode is fully linearly polarized 
(as the complete linear polarzation rules out any contribution from 
the second mode) and (b) leaves a deafening question regarding 
the fate of the second mode.
\item The trailing half of the pulsar's 21-cm profile exhibits a nearly 
equal contribution from the two polarization modes and is also very 
nearly depolarized.  The boundary between the highly polarized 
leading region and the negligibly polarized trailing region---shown 
near 0$\deg$ longitude in Figs.~\ref{fig10} \& \ref{fig12}---is precisely 
where the modulation-phase discontinuity is observed, arguing that 
the polarization and phase ``jumps'' are connected.  
\item The modulation-phase ``jump'' occurs in total-power analyses 
because the linear driftbands comprised of bi-polarization-mode 
subpulses cannot connect smoothly with those comprised of a single 
polarization mode.  

\end {itemize}

In summary, studies on {\em drifting} subpulses necessarily entail 
sightline cuts along the outside edges of conal beams, precisely the 
region now known to present the most complex (and interesting!) 
polarization-modal effects (\eg Rankin \& Ramachandran 2003).  If 
then modal polarization effects are the major cause of $\p2$ variations, 
it follows that modal effects must always be considered in measuring 
this fundamental parameter.

\acknowledgements 
We thank Russell Edwards  and Ben Stappers for interesting and 
useful discussions and their assistance with some of the Westerbork 
observations---and in particular their generously making available to 
us the results of their recent polarimetry calibrations.  We also thank 
Joeri van Leeuwen for assistance with some of the observations, 
Avinash Deshpande for various assistance, and Geoffrey Wright for 
discussions.  Portions of this work were carried out with support from 
the Netherlands Organisatie voor Wetenschappelijk Onderzoek and 
US National Science Foundation Grants AST 99-87654 and 00-98685.  
This work made use of the NASA ADS astronomical data system.

\end{document}